# Structure stability in the simple element sodium under pressure

V F Degtyareva[1] and O Degtyareva[2]

[1] Institute of Solid State Physics, Russian Academy of Sciences, Chernogolovka, 142432 Russia
[2] CSEC and School of Physics, University of Edinburgh, The King's Buildings, Mayfield Road, Edinburgh EH9 3JZ, UK

E-mail: degtyar@issp.ac.ru

**Abstract.**

The simple alkali metal Na, that crystallizes in a body-centred cubic structure at ambient pressure, exhibits a wealth of complex phases at extreme conditions as found by experimental studies. The analysis of the mechanism of stabilization of some of these phases, namely, the low-temperature Sm-type phase and the high-pressure $cI16$ and $oP8$ phases, shows that they satisfy the criteria for the Hume-Rothery mechanism. These phases appear to be stabilized due to a formation of numerous planes in a Brillouin-Jones zone in the vicinity of the Fermi sphere of Na, which leads to the reduction of the overall electronic energy. For the $oP8$ phase, this mechanism seems to be working if one assumes that Na becomes divalent metal at this density. The $oP8$ phase of Na is analysed in comparison with the MnP-type $oP8$ phases known in binary compounds, as well as in relation to the $hP4$ structure of the NiAs-type.

**Contents**

## 1. Introduction

The alkali metals are considered textbook examples of nearly-free-electron metals at ambient conditions. They crystallize in a highly symmetric and closely packed body-centred cubic (bcc) structure, with the highest Madelung constant of all metallic structures. On pressure increase, the bcc structure of alkali metals transforms to another close-packed structure, face-centred cubic (fcc) and then to unexpectedly complex low-symmetry structures (see reviews [1-4]). Lithium and sodium have been recently shown to adopt a complex cubic crystal structure containing 16 atoms in the unit cell (Pearson symbol *cI*16) at around 40 GPa [5] and at 105 GPa [4,6], respectively. On further compression of sodium, the experimentalists found that *cI*16 phase transforms at 118 GPa to an orthorhombic structure with 8 atoms in the unit cell (*oP*8) [7], a novel structure-type for an element. At 125 GPa, the *oP*8 phase transforms to an incommensurate host-guest structure [7,8] with the same 16-atom tetragonal host structure observed previously in K and Rb [9,10]. Melting line of Na showed a peculiar behaviour [11] with a bend-over at 31 GPa and 1000 K that reached its minimum of room temperature at 118 GPa. Numerous complex phases with large unit cells were discovered in the vicinity of the melting line minimum [7]. Heavier alkali metals, K, Rb, and Cs, have been long known to assume complex crystal structures under pressure above the stability region of fcc, at much lower pressures than in Li and Na. These include orthorhombic structures *oC*52 in Rb and *oC*84 in Cs, a host-guest structure in K and Rb, tetragonal *tI*4 and an orthorhombic *oC*16 structures in Rb and Cs (see reviews [1-4]). These phase transitions from fcc to complex structures in all alkali metals are accompanied by a decrease in symmetry, coordination number and packing density.

By computational and theoretical studies, the appearance of some of these complex structures in the heaviest alkali metals Rb and Cs is suggested to be connected to an electronic s to d transfer [12] (for recent review see [13]). On compression, the empty d electron level broadens and lowers in relation to the valence s-band, which results in a partial s to d electron transition. Similarly, the transition from fcc to complex phases at high pressures in Li and Na is explained by computational studies with an *s*-to-*p* band overlap [5,14]. The empty p electron level lowers in relation to the valence s-band and partial band overlap takes place. However, the observation of a large number of structural transformations and a variety of structural types in alkali metals under pressure require consideration of other factors that might be responsible for the formation of these phases. Hume-Rothery effect has been suggested as a stabilization mechanism for the *cI*16 structure in Li and Na, as well as for the *oC*52 and *oC*84 structures of Rb and Cs [3], that are related to *cI*16. This model explains the appearance of the complex structures as a result of lowering the electronic energy and thus the total energy of the crystal structure due to a formation of additional planes in a Brillouin-Jones zone in a close contact with the Fermi surface. This explanation has been developed for the so-called Hume-Rothery phases such as in the Cu-Zn alloy system long before the introduction of the band structure [15] and then verified by density-functional theory calculations [16]. The electronic contribution into total energy becomes more important on compression [17], adding significantly to the stabilization of the *cI*16 and related phases.

At higher pressures in the post-*cI*16 region, the density of Li and Na is so high ($\rho/\rho_0 \approx 4$ for Na) that the atomic cores are beginning to touch [18-20] and new physical phenomena come into play. To avoid the core overlap when calculating electronic structure at these densities, the 1*s* and 2*s* electrons of Li are treated as valence while in Na, 3*s*, 2*p*, and 2*s* electrons are considered valence and only the tightly bound 1*s* state is incorporated into an effective frozen core. These calculations find for both Na and Li that electrons localise in the interstitial regions of the structures [14,18,19,21-23]. Computational studies have suggested several crystal structures for the post-*cI*16 regime in Li and Na [14,18,19,21,22]. The crystal structures uncovered by experimental work, however, turned out to be quite different from the prediction, containing a much larger number of atoms in the unit cell [7,24].

Thus, the question about the factors governing the formation and stability of the high-pressure phases in Na still remains open. Here, we analyse some of the factors contributing to the stability of the *cI*16 structure in Na as well as in Li, and the *oP*8 structure in Na. We consider the Hume-Rothery mechanism as a driving force for the formation of these structures with the help of the recently developed computer program BRIZ for the visualization of the interaction of the Brillouin zones and Fermi sphere [25]. We compare the *oP*8 structure of Na to the MnP structure type known for many binary compounds and describe its relation to the *hP*4 structure of the NiAs-type. We also consider the martensitic transformation from bcc to a close-packed Sm-type structure (also known as 9R or hR3 in Pearson notation), that takes place at ambient pressure in both Li and Na by lowering temperature to 77 and 35 K, respectively [26,27]. Here we further explore the approach taken by Ashcroft [28] to describe the bcc to Sm-type transition within the Hume-Rothery mechanism.

## 2. Method of analysis

The energy of the crystal structure is known to consist of two main contributions: electrostatic energy and the band structure term. The latter usually favours the formation of superlattices and distorted structures. The energy of valence electrons is decreased due to a formation of Brillouin planes with a wave vector q near the Fermi level $k_F$ and an opening of the energy pseudogap on these planes if $q_{hkl} \approx 2k_F$. Within a nearly free-electron model the Fermi sphere radius is defined as $k_F = (3\pi^2 z/V)^{1/3}$, where z is the number of valence electrons per atom and V is the atomic volume. This effect, known as the Hume-Rothery mechanism (or electron concentration rule), was initially described to account for intermetallic phases [15], and then extended and widely used to explain the stability of complex phases in various systems, from elemental metals [3,28-31] to quasicrystals [32,33] and amorphous materials [34].

We analyze the mechanism of stabilization of some phases in Na using a computer program BRIZ [25] that has been recently developed to construct Brillouin zones or extended Brillouin-Jones zones (BZ) and to inscribe a Fermi sphere (FS) with the free-electron radius $k_F$. The resulting BZ polyhedron consists of numerous planes with relatively strong diffraction factor and accommodates well the FS. The volume of BZ's and Fermi spheres can be calculated within this program. The BZ filling by the electron states ($V_{FS}/V_{BZ}$) is estimated by the program, which is important for the understanding of electronic properties and stability of the given phase. For a classical Hume-Rothery phase $Cu_5Zn_8$, the BZ filling by electron states is equal to 93%, and is around this number for many other phases stabilized by the Hume-Rothery mechanism [25]. Diffraction patterns of these phases have a group of strong reflections with their $q_{hkl}$ lying near $2k_F$ and the BZ planes corresponding to these $q_{hkl}$ form a polyhedron that is very close to the FS. The FS would intersect the BZ planes if its radius, $k_F$, is slightly larger then $q_{hkl}/2$, and the program BRIZ can visualize this intersection. One should keep in mind that in reality the FS would usually be deformed due to the BZ-FS interaction and partially involved inside the BZ. The ratio $2k_F/q_{hkl}$, called a "truncation" factor or a closeness factor [35,36], is an important characteristic for a phase stabilized due to the Hume-Rothery mechanism. For Hume-Rothery phases such as $Cu_5Zn_8$, this closeness factor is equal 1.015, and it can reach up to 1.05 in some other phases. This means that the FS radius can be up to approximately 5% larger than the BZ vector $q_{hkl}/2$ for the phase stabilized due to a Hume-Rothery mechanism. Thus, with the BRIZ program one can obtain the qualitative picture and some quantitative characteristics on how a structure matches the criteria of the Hume-Rothery mechanism.

The stability of the high-pressure structures of Na is analysed here within the nearly-free electron model. This model is particularly valid in case of Na, as its FS is nearly spherical and remains spherical on compression in its bcc phase as well as in the fcc phase [19,20,37]. For the complex phases of Na, i.e. Sm-type, *cI*16, and *oP*8, an extended Brillouin-Jones zone is constructed with the planes that have a large structure factor, as seen in their diffraction patterns, and lie close to the $2k_F$. The matching criteria for the Hume-Rothery mechanism of the stabilization of these structures (i.e. the number of planes forming the BZ, the closeness factor, and the BZ filling by electron states) are then analysed.

## 3. Results
*3.1. Compact structures (bcc, fcc, Sm-type)*

First, we consider the compact structures experimentally observed in sodium (as well as in lithium), that are closely packed, have high coordination number and high Madelung constant, i.e. bcc, fcc and Sm-type (9R). Bcc has the highest Madelung constant $\alpha = 1.791858$, followed by fcc with $\alpha = 1.791747$ and hcp with $\alpha = 1.791676$ [38]. Thus, the stabilization of the bcc in alkali metals at ambient conditions can be understood as defined by the dominant electrostatic (or Madelung) contribution to the total energy of the crystal structure. The first-principle calculations show that all these close-packed structures have very similar energies with the difference less than 1-5 meV/ion, which about equal to the accuracy of the calculations [14,19,20]. This makes it impossible to say why one of these structures should be more favourable than the other. Thus, it seems worthwhile to look for some other factors that might contribute to the phase stabilization.

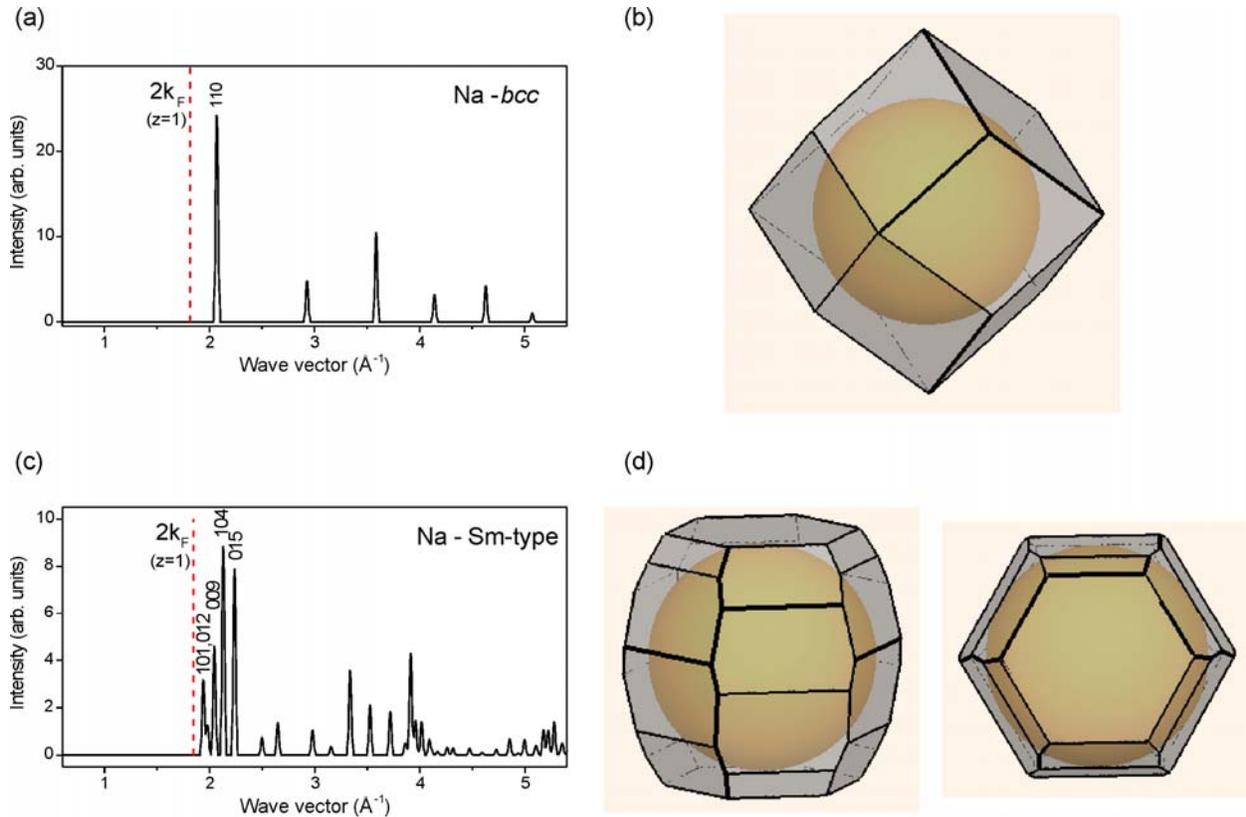

**Figure 1.** Calculated diffraction patterns of Na for (a) the ambient pressure bcc phase and (c) the Sm-type phase with structural parameters given in Table 1. (b,d) Corresponding Brillouin zones with the inscribed Fermi sphere. In (d), large Brillouin-Jones zone is shown in two different views, perpendicular and along the hexagonal c-axis. The position of $2k_F$ and the *hkl* indices of the planes used for the BZ construction are indicated on the spectra (for the rhombohedral Sm-type structure, the indices are given in hexagonal setting).

Let us consider the phase transition from bcc to Sm-type structure, experimentally found in both Li and Na at ambient pressure by lowering temperature to 77 and 35 K, respectively [26,27]. Stabilization of the Sm-type phase at low-temperature in Li has been previously attributed to the Hume-Rothery mechanism [28]. Because of a quite "spheroidal" shape of the Brillouin-Jones zone formed by numerous planes, "more of the lower-energy electrons are in states of higher density and the overall electronic energy is expected to be reduced" as Ashcroft puts it [28].

Here we extend this analysis further and construct a Brillouin-Jones zone for the Sm-type structure of Na. Its diffraction pattern contains a group of strong reflections close to $2k_F$, while bcc has only one 110 reflection in this q range (Fig. 1a and c). The BZ for Sm-type is formed by 26 planes, an increased number in comparison to 12 planes in bcc (Fig. 1b and d). The BZ for Sm-type phase is formed by 5 types of planes, three of which are closer to $2k_F$ than the 110 planes of bcc, as can be seen from Table 1 by comparing the $2k_F/q_{hkl}$ ratios for Sm-type and bcc. The BZ for Sm-type phase is also slightly more filled with electronic states (59%) than that for bcc (50%). Thus, as suggested for Li Sm-type in Ref. [28], in Na Sm-type, the overall electronic energy is expected to be reduced due to the Hume-Rothery mechanism, i.e. the formation of a BZ with numerous new planes with their vectors $q_{hkl}$ lying close to $2k_F$. We should note that at these conditions the electrostatic factor is very important for the structure stabilization in Na, and the possible choice of structures is among those close-packed, such as bcc, hcp, fcc and Sm-type. The latter is more favourable than bcc in terms of electronic energy due to Hume-Rothery mechanism and is stabilized at low-temperatures.

It is interesting to note that although FS of Li is nearly spherical at ambient pressure, the first-principle calculations report a significant distortion of FS in bcc Li already at 8 GPa, just before the transition to fcc [39]. Copper-like necks are found developed in fcc Li at about 30 GPa [39-41], so that the same Hume-Rothery mechanism is suggested for the stabilization of the fcc over bcc in Li [39]. Indeed, the planes (111) of the fcc BZ (eight planes in total) are closer to $2k_F$ than (101) planes of bcc (12 planes in total). The corresponding $2k_F/q_{hkl}$ ratio is slightly closer to 1 in fcc compared to bcc (Table 1), which could lead to an increased interaction between the FS and the BZ planes making fcc more favourable structure. Na, however, is calculated to have an approximately spherical FS in the bcc phase [20], as well as in the fcc phase [37], stable above 65 GPa [42], which does not speak in favour of Hume-Rothery mechanism [43]. The bcc to fcc transition in Na has been characterized as a tetragonal Bain path with the driving force attributed to the changes in lattice dynamics and elastic constants [43]. The electronic properties, however, can be significant in driving tetragonal Bain path as was demonstrated in Ref. [44].

*3.2. Na-cI16, distortion of bcc*

Next, we analyse the stability of the complex low-symmetry low-coordination structures, found in Na above 104 GPa. In this work, we particularly focus on the *cI*16 and the *oP*8 structures. The *cI*16 structure has been discovered experimentally in Li at around 40 GPa [5] and then consequently in sodium stable in the pressure range from 105 to 117 GPa [4,6,7]. The transition from fcc to the *cI*16 structure is accompanied by the lowering of the coordination number from 12 to 11 and by the reduction of the nearest neighbour distance and packing density (Fig. 2). This structure is stabilized by factors other than the minimization of electrostatic energy which favours symmetric and close-packed arrangement of ions. We are going to look at the reduction of the electronic energy as a stabilization factor for the *cI*16 phase.

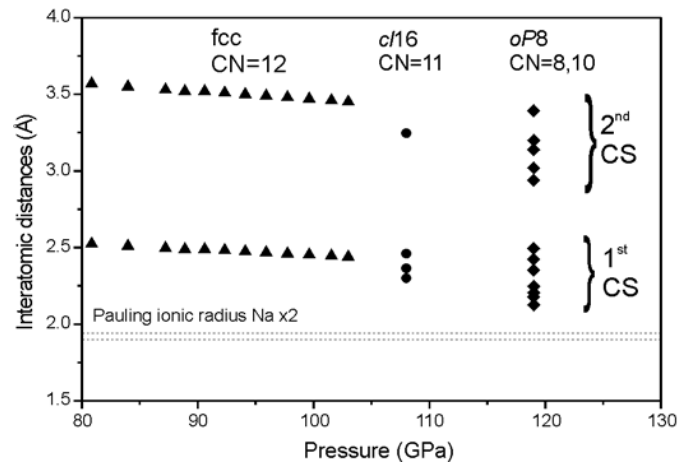

**Figure 2**. Pressure dependence of interatomic distances of Na for the fcc, *cI*16 and *oP*8 phases. "CN" stands for coordination number. The 1st and 2nd coordination shells (CS) are outlined. Data points for fcc are calculated using the equation of state reported in [42], values for the *cI*16 and *oP*8 phases are calculated using the parameters listed in Table 1. The value of Pualing ionic radius is shown, known for Na at ambient pressure.

This structure can be considered as a 2 x 2 x 2 supercell of bcc with a small displacement of the positions of the atoms along the space diagonal. As discussed earlier in Refs. [3,29,30], it is this displacement that leads to the formation of new diffraction reflections observed in the x-ray diffraction pattern compared to bcc, and importantly, the reciprocal lattice vector for the first diffraction reflection (211) lies in the vicinity of the $2k_F$ (Fig. 3a). This means that the planes (211) of the Brillouin-Jones zone lie almost on the Fermi sphere ($2k_F/q_{211}$ is very close to 1 as can be seen in Table 1). Here we construct the Brillouin-Jones zone for the *cI*16 structure of Na. It is built of the (211) planes and its general view is shown in Fig. 3b. This high-symmetry polyhedron with 24 planes accommodates the Fermi sphere very well with an electron state filling (in the free electron model) of roughly 90%. The criteria of the Hume-Rothery stabilization mechanism are satisfied.

The analysis of the BZ and FS configuration of the *cI*16 structure suggests that the reason for the stabilization of this low-symmetry structure is the Hume-Rothery mechanism. The energy of this phase lowers when the new planes of the Brillouin zone form near the Fermi level and an energy gap opens on these planes. The Fermi sphere - Brillouin zone interaction modifies the electron density of states with a substantial rise near (below) $k_F$ and reduction directly at $k_F$. This qualitative picture is in tune with the first-principle calculation of electronic density of states for Li-*cI*16 [5]. Calculations of the Fermi surface of bcc Li in the cI16 stability field show that it becomes increasingly anisotropic with pressure and develops an extended nesting along the bcc [211] direction [45], which goes along with our model shown in Fig. 3b. In our analysis we made one step further to actually *draw* the new BZ planes that appear in the *cI*16 structure in comparison with bcc, to show that they lie near the Fermi surface casing its deformation and reduction of electronic energy. Thus, the special stability of the *cI*16 structure in Li as well as in Na is due to the Fermi surface contacting all faces of the Brillouin-Jones zone with electronic energy significantly lowered due to the gap across the zone faces.

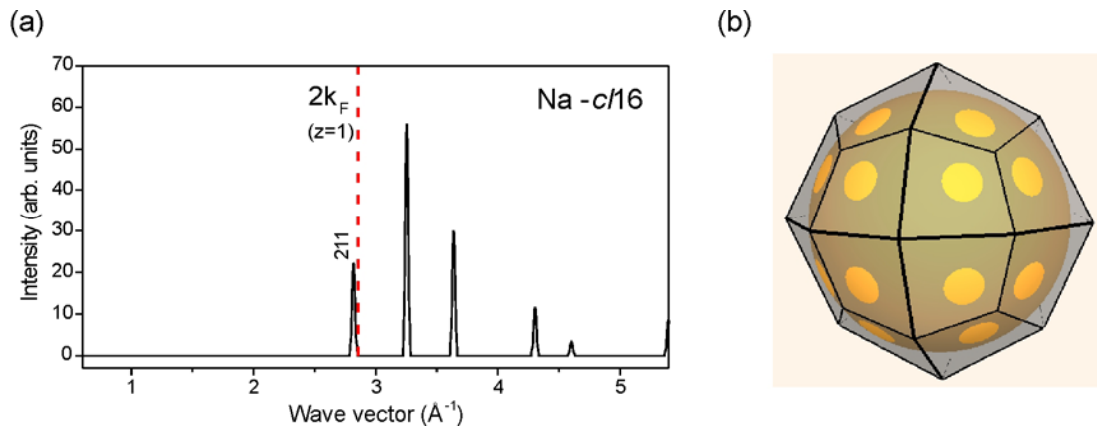

**Figure 3.** (a) Calculated diffraction pattern for the high-pressure phase of Na, *cI*16, and (b) corresponding Brillouin zone with the inscribed Fermi sphere. The position of $2k_F$ and the *hkl* indices of the plane used for the BZ construction are indicated in (a).

*3.3. Na-oP8, distortion of NiAs-hP4*

*3.2.1. oP8 – a Hume-Rothery phase.* Now using the same concept, we analyse the stability of the next high-pressure phase of sodium with the orthorhombic *oP*8 structure, stable between 117 and 125 GPa [7]. We use the first group of strong diffraction reflections observed in the x-ray diffraction pattern of Na-*oP*8, (200), (102) and (111) (*italic hkl* indices in Fig. 4a), lying close to $2k_F$ (z=1), to construct the first Brillouin-Jones zone (not shown). Estimation of the filling of this zone with the electronic states per atom gives a value of 97% which is too high in comparison with 89-93% zone filling typical for the Hume-Rothery phases. Also the ratio of $2k_F/q_{hkl}$ (the closeness factor) is too high (1.07-1.10) in comparison with the value of 1.015 obtained for the Hume-Rothery phase $Cu_5Zn_8$ [25]. These conditions do not satisfy the criteria for the Hume-Rothery mechanism. This analysis is done assuming the number of valence electrons per atom equal 1. In what follows we show that Na needs to be assumed to be a divalent metal ($k_F$ is calculated assuming $z = 2$), in which case the Na $oP8$ phase satisfies the criteria of the Hume-Rothery mechanism.

The $oP8$ structure found in Na is unique among the elements. This structure belongs to the MnP structure type (the *Stukturbericht* symbol is B31) and is known for as many as 30 binary phases [46]. There is a close similarity in atomic positions of the space group *Pnma* and in the axial ratio between those phases and Na-$oP8$ (see below). Thus, Na-$oP8$ is *isostructural* with the MnP structure type. The binary phases of MnP-type consist of one transition metal and one *sp* element from groups III to VI. There is however one example of the $oP8$ phase which consists of two simple metals, i.e. AuGa. There is also another example of a simple metal binary phase – AuAl – which is only a small monoclinic distortion of the $oP8$ structure (an $mP8$ structure with a monoclinic angle of 93°) (Table 2). The $oP8$ structures of Na and AuGa and the $mP8$ structure of AuAl have remarkably close axial ratios and atomic positions (Table 2).

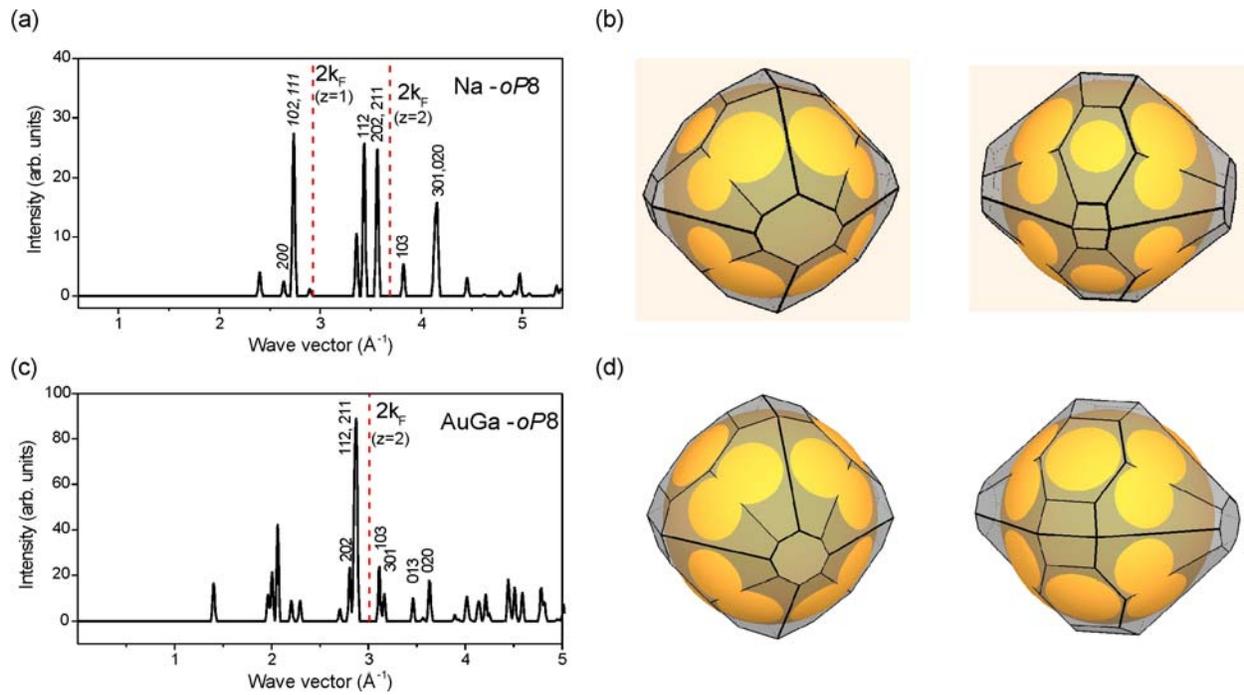

**Figure 4.** Calculated diffraction patterns for (a) the high-pressure phase of Na, $oP8$, and (c) the ambient pressure $oP8$ phase of AuGa. (b,d) Corresponding Brillouin zone with the inscribed Fermi sphere in projections close to [010] (left) and [100] (right). The position of $2k_F$ for $z = 1$ and 2 for Na and $z = 2$ for AuGa, calculated from a free-electron model, is shown in (a) and (c), respectively. The non-italic hkl indices indicated in (a) and (c) correspond to the planes used for the BZ construction.

The *sp* metal phases AuGa and AuAl have an average number of valence electrons per atom equal 2 and give us an opportunity to apply a nearly-free-electron model to the analysis of their structural stability within the Hume-Rothery mechanism. Following the analysis of the complex binary phases presented in Refs [25,36], one can show that the $oP8$ phase in AuGa and the $mP8$ phase in AuAl satisfy the Hume-Rothery stabilization criteria. Indeed, the *strongest* group of reflections in the diffraction pattern (Fig. 4c) lies close to $2k_F$, which is calculated assuming $z = 2$ (one valence electron from Au and three valence electrons from Ga give an average number of valence electrons per atom equal 2). Corresponding Brillouin-Jones zone contains numerous planes (Fig. 4d), lies close to the Fermi sphere, and is filled with electronic states by 94%. Thus, we have shown that the $oP8$ phase in AuGa is stabilized due to a Hume-Rothery mechanism. The electronic factor of stabilization is not so evident when a transition metal participates in this phase, nevertheless one can infer that 2 electrons per atom is a necessary condition of stability for this complex phase and this condition can be extended to Na-$oP8$.

Now for Na-$oP8$, we select the next group of strong reflections on the diffraction pattern (Fig. 4a), that lies close to $2k_F$ calculated from a free-electron model assuming the number of valence electrons per atom $z = 2$. The constructed BZ (Fig. 4b) consists of 20 planes (112), (202) and (211) that are in contact with the FS. These planes together with (103), (301) and (020) form a polyhedron that accommodates very well the free-electron FS with some intersections. Zone filling by electron states is ~93% - the same value as in the case of the Hume-Rothery phase $Cu_5Zn_8$ [25]. This and the closeness factor (Table 1) satisfy very well the criteria for the Hume-Rothery mechanism.

The assumption of $z = 2$, i.e. Na becoming a divalent metal in the $oP8$ phase, means that some of the core electronic bands (partially) move close to $k_F$. From the electronic band structure calculations, the participation of core electrons in the valence band has been suggested for Na at pressure of 130 GPa close to the stability region of $oP8$ [19]. Also, at the densities of Na-oP8 (approx. four-fold compression), the ion cores begin to overlap. "Hence, the concept of well determined ion cores, which is the basis of standard metallic bond descriptions, becomes inapplicable at ultrahigh pressures" [20].

At the transition from $cI16$ to $oP8$ structure in Na, the coordination number further decreases from 11 to 8 and 10 for two different atomic sites (Fig. 2). At 119 GPa, the interatomic distances lie in the ranges 2.125-2.493 Å (for one atomic site) and 2.125-2.253 Å (for the other atomic site). The shortest distance corresponds to the atomic radius of 1.062 Å, that is very close to the Pauling ionic radius of $Na^+$ equal 0.95(0.97) Å [46] (Fig. 2). To attain this atomic arrangement, Na electronic structure would need to change drastically in comparison with the lower-pressure fcc and $cI16$ phases considerably reducing the atomic radius (our estimation give approximately 4%). As proposed in the present work, this can be achieved by the partial "ionisation" of the Na core when one electron is moved from the core into valence band, thus Na becoming a divalent metal.

In this respect, it is interesting to look at the alkali metals being involved in a formation of compounds with transition metals. In such compounds, the alkali metals are subjected to "chemical pressure", as the ambient pressure atomic volume of alkali metals is much larger than that of transition metals. Such "chemical pressure" can be rather significant and comparable to the external pressure and produce changes in the valence electron configuration, such as participation of core electrons in the valence band. This assumption is supported by the formation of compounds that are defined by a particular electron concentration due to a Hume-Rothery mechanism. Hume-Rothery, considering the $Ag_5Li_8$ γ-brass phase in his textbook [47], suggested that its stability could be also connected to the electron concentration 21/13 as for other γ-brass phases, if lithium were divalent. Other examples include the ambient pressure compounds $Li_2Pd$ and $Li_2Pt$ [48] and a high pressure phase $K_2Ag$ [49] that have a hexagonal structure $hP3$ similar to the so-called δ-phase formed in the Cu-Zn alloy system [48] that belongs to the family of Hume-Rothery phases. The stabilization of these phases due to the Hume-Rothery mechanism would require the alkali metals in these phases to be in the divalent state. This effect occurs in some other alkali-transition metal compounds where the electron concentration factor is not so evident competing with other factors that control structure stability.

*3.2.2. Relation to NiAs-hP4.* The $oP8$ MnP-type structure is a distortion of the $hP4$ NiAs-type as can be seen from Fig. 5a showing an atomic network in projection along the [010] direction of the $oP8$ which corresponds to the [100] direction of the $hP4$ structure. Relation of orthorhombic (o) and hexagonal (h) lattices is as follows: $a_o \approx c_h$, $b_o \approx a_h$, and $c_o \approx a_h\sqrt{3}$. In the $oP8$ structure, atoms are slightly displaced from the ideal positions of the $hP4$ structure. For 20 binary phases with $oP8$ MnP-type structure, the atomic positions lie within the narrow ranges: for two atomic positions (4c) of the *Pnma* space group $x_1 = 0.005 - 0.010$, $z_1 = 0.182 - 0.202$ and $x_2 = 0.182 - 0.190$, $z_2 = 0.564 - 0.590$, as summarized in Ref. [50]. The atomic positions of Na-$oP8$ lie close to these values. The structural transformation from $hP4$ NiAs-type to $oP8$ MnP-type is accompanied by an increase in coordination number from 6 in $hP4$ to 10 and 8 for two different sites of $oP8$.

The formation of the distortive *oP*8 structure with certain atomic displacements from *hP*4 leads to the appearance of numerous additional diffraction peaks (Fig. 4a) compared to the simple diffraction pattern of the *hP*4 phase (Fig. 5b). The (102) reflection of the *hP*4 structure that lies close to $2k_F$ (calculated assuming z = 2) (Fig. 5b) splits into (202) and (211) in the *oP*8 structure and close to them an additional strong reflection (112) appears due to the structural distortion. From the point of view of the Brillouin-Jones zone, in *oP*8 a new additional plane (112) is added to the BZ (Fig. 4b) compared to *hP*4 reducing the volume of the BZ, which leads to an increase in the BZ-FS interaction and reduction of the electronic energy.

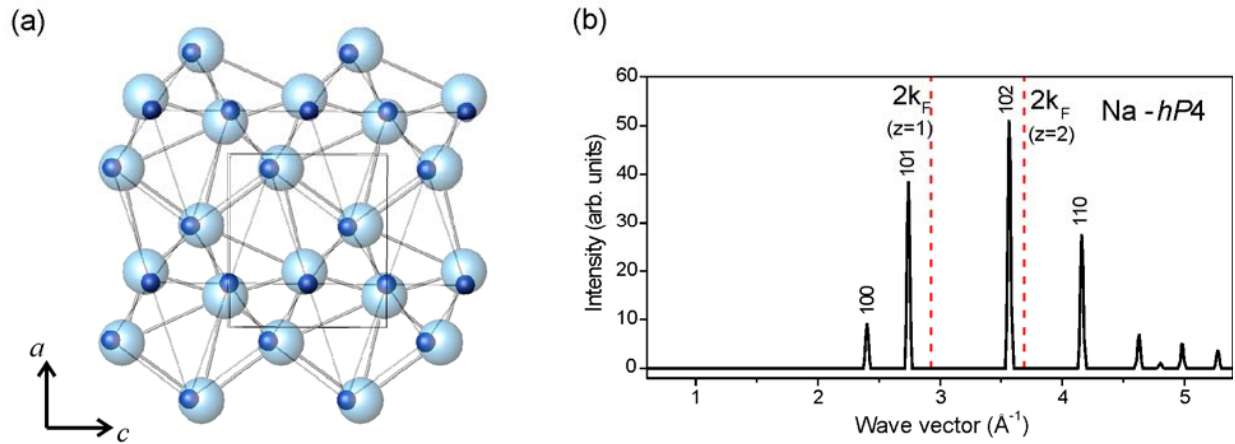

**Figure 5**. (a) Structural deformation of the *hP*4 structure (dark blue atoms) into *oP*8 structure (light blue atoms). For the *oP*8 structure, the lattice parameters and atomic positions of Na measured at 119 GPa [Greg08] are used (see Table 1). For the *hP*4, the following atomic positions were taken (in orthorhombic setting): 4c (¼ ¼ 0.583) and 4c (0 ¼ ¼), with the same lattice parameters as in Na-*oP*8. The atomic bonds up to 3.67 Å are shown. All atomic displacements are in the plane of drawing. The layer at y = ¼ is shown, the other layer at y = 3/4 can be obtained by rotating this layer by 180° around the a-axis and 180° around the c-axis. (b) Calculated diffraction patterns for a hypothetical *hP*4 structure of Na with the structural parameters as in (a) (in hexagonal setting: space group $P6_3/mmc$, a = 3.020 Å, c = 4.765 Å, Na is in 2a position (0 0 0) and 2c position (1/3 2/3 ¼)). The hkl indices are given in hexagonal setting. The position of $2k_F$ for z = 1 and 2 for Na is shown.

Phase transitions between *hP*4 and *oP*8 phases are known as a function of temperature and composition, as considered in [Franzen PRB 74]. The studies involving phases with the NiAs structure (*hP*4) indicate that "it tends to be stable at high temperature and transform to other structures at lower temperature" (pp. 454-456 in Ref. [46]), as for example in the CoAs alloy [48]. Another example of the transformation NiAs-type (*hP*4) to MnP-type (*oP*8) is the FeS alloy. The case of this alloy is especially interesting because the MnP-*oP*8 phase is stabilised at high pressure [52-54]. There are more examples demonstrating the phase transformation under pressure from NiAs-*hP*4 to MnP-*oP*8: MnAs [55,56], MnTe [57], CrSb [58], CrTe [59]. This transition is accompanied by a change in magnetic properties associated with the change in the electronic structure and atomic rearrangement. In the case of non-magnetic elements, the *hP*4 to *oP*8 transition is driven by the reduction of valence electron energy. This is especially evident in the case of an alloy with simple metals as constituent elements, i.e. AuGa-*oP*8. AuSn is an example of *hP*4 structure in a simple metal alloy.

The relation of the *oP*8 structure to the *hP*4 NiAs type can be extended to a wider structural group of hexagonal phases with the common *Structurbericht* type B8 that combines Ni$_2$In and NiAs structure types; the former is a filled up type of the latter. The number of the known B8-type phases has been growing steadily to a today total of more than 400 compounds. "The B8 structure type is unique in that it offers such a rich flora of examples of diverse structural behaviour" (superstructures, ordering of vacancies etc) [60]. Lower-symmetry variants, such as orthorhombically distorted MnP (*oP*8) as well as Co$_2$Si (*oP*12) structures, are the distortions of the hexagonal structures of NiAs-type (*hP*4) and Ni$_2$In-type (*hP*6), respectively, and are of special interest in relation to Na-*oP*8.

*3.2.3. Electron configuration in Na-oP8.* The *oP*8 structure found in Na has no structural analogous among elements. A search for some binary analogues yielded in comparison of this structure to an anti-cotunnite structure [61] (PbCl$_2$ is a prototype of cotunnite and a representative example of anti-cotunnite is Co$_2$Si). Na would occupy the cation atomic positions omitting the anion positions, where a localisation of the electron charge is suggested. This structure was found, for example, in a high-pressure form of Li$_2$O [62], that has the same space group *Pnma* and axial ratios similar to Na-*oP*8 This interpretation of the Na-*oP*8 structure is within the concept of relating high-pressure forms of alkali metals to certain compounds, so-called electride structures, with alkali metal atoms occupying the cation positions and electron density maxima located in the empty anion positions [2,63,64]. For example, in the Cs-IV (*tI*4) structure, caesium atoms are suggested to occupy thorium positions in the structure of the binary ThSi$_2$ phase, viewing this high-pressure modification of Cs as cesium electride [63]. Similarly, the host-guest structure of Rb-IV is compared with an intermetallic phase W$_5$Si$_3$ and orthorhombic Cmca structure of Cs-V with a ternary compound KLiO, again suggesting that the alkali metal atoms occupy the atomic positions of cation atoms and proposing an electron change maxima in the anion positions [2]. In tune with this approach, the theoretical calculations for light alkali elements Li and Na have suggested that at strong compression electrons are forced away from the near core regions resulting in an increased electron density in the *interstitial* regions [19,65].

As shown above, the Na-*oP*8 structure is crystallographically related to the family of the *Structurbericht* B8 type containing the Ni$_2$In-type (*hP*6) and NiAs-type (*hP*4) structures through an orthorhombic distortion to the Co$_2$Si-type (*oP*12) and MnP-type (*oP*8) structures. The interpretation of the Na-*oP*8 phase as an "electride" relates it to the Co$_2$Si (*oP*12) structure with Na occupying the cation atomic positions with the interstitially localized electrons occupying empty anion positions. In present work, an alternative approach is suggested for the interpretation of Na-*oP*8, in which Na atoms occupy the positions of the AuGa structure of the MnP-type (*oP*8) that is considered as a Hume-Rothery phase stabilized with two valence electrons per atoms. This interpretation has led us to the assumption of Na-*oP*8 becoming a divalent metal.

## 4. Summary and conclusions

The compressed alkali metals assume complex low-symmetry structures that are quite unusual among elements. The question arises as to what extent these structures are idiosyncratic, how do they relate to the known structural types and whether they follow any known classification (see, for example, Ref. [46]). It is useful to look for relations between the structures of compressed alkali metals and some binary phases and to consider main factors responsible for the formation of complex phases in simple metals and alloys. Here we have shown that the *oP*8 structure of Na is isostructural with the MnP-type and has a close similarity with the binary AuGa *oP*8 phase. Assuming the Hume-Rothery mechanism to be the main factor for the stabilization of *oP*8 in AuGa and Na, we suggest that at the densities of the *oP*8 phase of Na core electrons start to participate in the valence band and Na becomes a divalent metal.

The *oP*8 structure of Na is analysed here as a distortion of an *hP*4 structure of NiAs-type. We have shown that for a simple metal, the *oP*8 structure would have a reduced electronic energy in comparison with the *hP*4 structure. For the alkali metals Li, Na and K, one could expect to find a possible phase transition between the *hP*4 and *oP*8 structures, as it is known to occur in many binary compounds with change in pressure, temperature or composition.

The high-pressure *cI*16 phase of Na is shown to be stabilized due to the Hume-Rothery mechanism, similar to the *cI*16 structure of Li. Entirely new planes are form in the Brillouin zone in the vicinity of the one-electron Fermi surface, providing a reduction of the electronic energy. For the low-temperature close-packed Sm-type structure of Na, a reduction of the electronic energy in comparison to bcc is proposed due to a formation of an almost spheroid Brillouin zone.

**Acknowledgments**
Financial support by the Russian Foundation for Basic Research through grant 02-07-00901 is gratefully acknowledged. O.D. acknowledges support from the Royal Society.

**Table 1.** Structure parameters of Na phases as given in the literature. The Fermi sphere radius $k_F$, the total number of Brillouin zone planes, the ratio of $2k_F$ to Brillouin zone vectors ($2k_F/q_{hkl}$) and the degree of filling of Brillouin zones by electron states $V_{FS}/V_{BZ}$ are calculated by the program BRIZ [25].

| Phase | Na-bcc | Na Sm-type | Na-fcc | Na-*cI*16 | Na-*oP*8 |
|---|---|---|---|---|---|
| **Structural data** | | | | | |
| **Pearson symbol** | *cI*2 | *hR*9 | *cF*4 | *cI*16 | *oP*8 |
| **Space group** | *Im*-3*m* | *R*-3*m* | *Fm*-3*m* | $I\bar{4}3d$ | *Pnma* |
| **P,T conditions** | Ambient conditions | Ambient pressure, T = 20K | P=65.1 GPa T=300 K | P=108GPa T=300K | P=119GPa T=300K |
| **Lattice parameters (Å)** | a = 4.2908 | hex. axes a = 3.766 c = 27.653 rhomb. axes a = 7.807 a = 22.93° | a = 3.6627 | a = 5.461 | a = 4.765 b =3.020 c = 5.251 |
| **Atomic volume (Å³)** $V/V_0$ | 39.50 1 | 37.74 0.955 | 12.28 0.311 | 10.18 0.258 | 9.45 0.239 |
| **Atomic positions** | 2*a* (0 0 0) | hex. axes 3*a* (0 0 0) 6*c* (0 0 ²/₉) rhomb. axes 1*a* (0 0 0) 2*c* (²/₉ ²/₉ ²/₉) | 4*a* (0 0 0) | 16*c* (*x,x,x*) *x* = 0.044(1) | 4*c* ($x_1$,¼,$z_1$) 4*c* ($x_2$,¼,$z_2$) $x_1$ = 0.015 $z_1$ = 0.180 $x_2$ = 0.164 $z_2$ = 0.586 |
| **Reference** | [66] | [27] | [42] | [6] | [7] |
| **FS - BZ data from the BRIZ program** | | | | | |
| **z (number of valence electrons per atom)** | 1 | 1 | 1 | 1 | 2 |
| $k_F$ (Å⁻¹) | 0.908 | 0.922 | 1.341 | 1.427 | 1.844 |
| **Total number of BZ planes** | 12 | 26 | 14 | 24 | 30 |
| **HKL and $2k_F/q_{hkl}$** | *110*: 0.877 | hex. axes: *101*: 0.951 *012*: 0.932 *009*: 0.902 *104*: 0.866 *015*: 0.825 | *111*: 0.903 *200*: 0.782 | *211*: 1.013 | *112*: 1.074 *202*: 1.035 *211*: 1.034 *103*: 0.964 *301*: 0.892 *020*: 0.887 |
| **Filling of BZ with electron states $V_{FS}/V_{BZ}$ (%)** | 50% | 59% | 50% | 89% | 93% |

**Table 2.** Structural parameters of the *oP*8 phase of Na [7] in comparison with the *oP*8 phase of AuGa and the *mP*8 phase of AuAl.

| Phase | Na-oP8 | AuGa | AuAl |
|---|---|---|---|
| **Pearson symbol** | *oP*8 | *oP*8 | *mP*8 |
| **Space group** | *Pnma* | *Pnma* | *P2$_1$/m* |
| **P,T conditions** | P=119GPa T=300K | Ambient conditions | Ambient conditions |
| **Lattice parameters (Å)** | $a$ = 4.765 $b$ = 3.020 $c$ = 5.251 | $a$ = 6.267 $b$ = 3.461 $c$ = 6.397 | $a$ = 6.339 $b$ = 3.331 $c$ = 6.415 $\beta$ = 93.04° |
| **Axial ratios** | $a/b$ = 1.578 $c/a$ = 1.102 $c/b$ = 1.739 | $a/b$ = 1.811 $c/a$ = 1.021 $c/b$ = 1.848 | $a/b$ = 1.903 $c/a$ = 1.012 $c/b$ = 1.845 |
| **Atomic positions** | 4*c* (0.015 ¼ 0.180) 4*c* (0.164 ¼ 0.586) | Au 4*c* (0.010 ¼ 0.184) Ga 4*c* (0.195 ¼ 0.590) | Au1 2*e* (0.189 ¼ 0.026) Au2 2*e* (0.313 ¼ 0. 500) Al2 2*e* (0.576 ¼ 0.195) Al1 2*e* (0.964 ¼ 0.695) |
| **Reference** | [7] | [67] | [68] |